\documentclass{raa}            

\usepackage{graphicx,times}             

\begin{document}

   \title{Formation of a rapidly rotating classical Be-star in a massive close binary system
}

   \volnopage{Vol.0 (20xx) No.0, 000--000}      
   \setcounter{page}{1}          

   \author{Evgeny Staritsin
   }

   \institute{K.A. Barkhatova Kourovka Astronomical Observatory, B.N. Yeltsin Ural Federal University,
             pr. Lenina 51, Ekaterinburg 620000, Russia; {\it Evgeny.Staritsin@urfu.ru}\\
   }

   \date{Received~~2009 month day; accepted~~2009~~month day}

\abstract{
This paper investigates the
spin-up of
a
mass-accreting star in a close binary system passing through the first stage of mass exchange in the Hertzsprung gap.
Inside an
accreting star,
angular momentum is carried by meridional circulation and shear turbulence.
The circulation carries part of the angular momentum entering the accretor to its surface.
The greater the rate of arrival of angular momentum in the accretor is, the greater this part.
It is assumed that this part of the angular momentum can be
removed by the disk
further
from the accretor.
If the angular momentum
in the matter entering the accretor is
more
than half
the Keplerian value,
then
the angular momentum obtained by the accretor
during mass exchange stage
does not depend on the rate of arrival of
angular momentum. The accretor may have the characteristics of a
Be-star immediately after the end of mass exchange.
}

   \authorrunning{E. Staritsin }            
   \titlerunning{Formation of a rapidly rotating classical Be-star }  

   \maketitle

%
%
\section{Introduction}           
\label{sect:intro}

Classical Be-stars include OBA stars with observed or previously observed
emissions
in the Balmer lines of hydrogen (Porter \& Rivinius \cite{PR03}). These stars are not
supergiants
and have large rotational velocities.
Among Be-stars, there is the group
of
early spectral subclasses (B3-O9).
The surface rotational velocities of these stars
range widely.
The lower range limit is 40\%-60\% of the Keplerian value,
while
the upper limit is 90\%-100\% (Cranmer \cite{C05}). The origin of the large rotational velocities of Be-stars is not clear.

Young B-stars
in the
early spectral subclasses and O-stars are characterized by lower rotational velocities (Huang et al. \cite{HGMS2010}). 70\% of these stars are observed in binary and multiple systems (Chini et al. \cite{CHN2012}; Sana et al. \cite{SMK2012}). All these stars are expected to form binary and multiple systems, considering
selection effects. Mass exchange in a binary system may be the reason for the rapid rotation of
the
star receiving mass.
\pagebreak[4]
The synthesis of
the
Be-stars
population in binary systems makes it possible to reproduce the observed number of these stars in the Galaxy (Pols et al. \cite{PCW91}; Portegies Zwart \cite{PZ95}; Van Bever \& Vanbeveren \cite{VV97}; Shao \& Li \cite{SL14}; Hastings et al. \cite{HLW21}).

A simple estimation made assuming the instantaneous redistribution of angular momentum in the stellar interior to
solid-state rotation shows that a 5\% -- 10\% increase
to
the star's mass due to accretion of mass with
Keplerian velocity leads
to a critical rotation state (Packet \cite{P81}).
The question of what happens when there is continued accretion into a star close to a state of critical rotation has been discussed in Paczy{$\acute{\mbox{n}}$}ski (\cite{P91}),
Popham \& Narayan (\cite{PN91}), Colpi et al. (\cite{CNC91}), and Bisnovatyi-Kogan (\cite{BK93}). Paczy{$\acute{\mbox{n}}$}ski (\cite{P91}), Popham and Narayan (\cite{PN91}),
and Colpi et al. (\cite{CNC91}) used various approaches. All authors agree that accretion does not stop when the star's speed of rotation reaches a critical value.
Paczy{$\acute{\mbox{n}}$}ski (\cite{P91}) studied the whole star-boundary layer-accretion disk system for various rotations of the central star. For models rotating slightly above critical, mass accretion is accompanied by the loss of angular momentum from the star to the disk, mediated by viscous stresses. However, the solutions obtained in Paczy{$\acute{\mbox{n}}$}ski (\cite{P91}), Popham and Narayan (\cite{PN91}), and Colpi et al. (\cite{CNC91}) are not self-consistent. The condition for a self-consistent solution for a system consisting of a star
in a state of critical rotation and an accretion disk is that "the star absorbs accreted matter with a certain angular momentum, such that the star remains in a state of critical rotation"
(Bisnovaty-Kogan \cite{BK93}). Let $J(M)$ be the angular momentum of a star with mass $M$ in a state of critical rotation and let $j_e^{Kep}$ be the specific Keplerian angular momentum
at the equator of the star. Then $j_a=dJ/dM<j_e^{Kep}$. A mass-accreting star can move along the sequence of stars in a state of critical rotation $J(M)$ if the excess angular momentum
of $\bigtriangleup j=j_e^{Kep}\!-\!j_a$ is eliminated. Bisnovaty-Kogan (\cite{BK93}) constructed models of accretion disks that remove excess angular momentum from the surface of a star.
At the same time, the speed of rotation at the star's surface remains critical.
So an increase in the mass and angular momentum of a star in a critical rotation state may occur due to the removal
of
excess angular momentum from the star by the accretion disk (Paczy{$\acute{\mbox{n}}$}ski \cite{P91}; Bisnovatyi-Kogan \cite{BK93}).

Physical processes, such as meridional circulation and turbulence, require finite
amounts of
time to transfer
angular momentum (Staritsin \cite{St19}, \cite{St21}). At the very beginning of accretion, only the outer layers of the star, including the accreted mass, have a fast rotation. The star surface
gains
critical rotation shortly after the start of accretion. Later, at the accretion stage in a state of accretor critical rotation, the circulation carries part of the angular momentum
brought along
with the accreted mass from
subsurface layers to the
star's
surface
(Staritsin \cite{St2022}). Thus, accreted layers can shrink, as
usually happens during accretion. The angular momentum transferred by circulation to the
star's
surface
can be
removed
through an accretion disk (Paczy{$\acute{\mbox{n}}$}ski \cite{P91}; Bisnovatyi-Kogan \cite{BK93}). Thus, the mass and angular momentum of an accretor in a state of critical rotation increase due to the removal of excess angular momentum from the accreted layers to the accretor surface and the further removal of this angular momentum from the star.

In
Staritsin (\cite{St2022}), the transfer of angular momentum in the accretor interior was carried out only by meridional circulation. The turbulence was artificially suppressed. This made it possible to elucidate the transport properties of circulation in
an accreting
star's interior.
The role of turbulence in
angular momentum transport within the accretor remained unclear. As to the angular momentum input, only one option has been considered,
the
effective transport of angular momentum from the
disk's
boundary layer
to the
accretor's
upper layer.

In this paper, we consider two mechanisms of angular momentum transfer in
an accreting star interior, namely circulation and turbulence. This allows us to find
the role of turbulence in the
spinning up
of a star.
We also took into account the possible reduction of
input angular momentum. This decrease can be attributed both to the transfer of angular momentum from the boundary layer to the outer parts of the disk, and to
sub-Keplerian rotation in the disk. The
accretor's
rotation, obtained
as a result of mass exchange, has been studied depending on the angular momentum introduced
during
mass exchange.

\section{Basic equations and simplifications}
\label{sect:basic}

\subsection{The angular momentum input}
\label{subsect:increment}

The matter lost by the donor due to the filling of the Roche lobe falls
into
the
accretor's
gravitational field
and swirls around it. The formation of gas structures
around the accretor, in particular the formation of
a
disk and
a
velocity field in it, depends on the ratio between three
factors
: the size of the accretor $R$, the minimum distance $\omega_{min}$ from the center of the accretor to the central line of the stream of matter falling from donor point $L_1$, and the distance from the
accretor's
center
to the edge of the inviscid disk $\omega_d$ (Lubow \& Shu \cite{LS75}).

Transient disks with sub-Keplerian rotation
have been
found (for example, RW Tau (Kaitchuck \& Honeycutt \cite{KH1982}) and $\beta$ Per (Cugier \& Molaro \cite{CM1984}, Richards \cite{R1992})) in
direct-impact systems ($\omega_d<R$). Three-dimensional hydrodynamic calculations show
disk
formation
in such systems;
the rotation velocity is 80\% and 60\% of the Keplerian value at the inner and outer edges of the disk, respectively
(Raymer \cite{R2012}).

Both transient disks  (SW Cyg) and permanent, but  variable, accretion disks  (RY Gem, TT Hya, AU Mon) in
grazing-impact systems ($\omega_{min}<R<\omega_d$)
have been discovered.
The velocity fields in the transient disk of
the
SW Cyg system and in the permanent disk of
the
RY Gem system are sub-Keplerian (Kaitchuck \cite{K1988}, \cite{K1989}). Asymmetric parts were found in the disks
of the
TT Hya and AU Mon systems;
the gas in the disk's asymmetric part in
the AU Mon system moves
at a
sub-Keplerian velocity (Richards et al. \cite{RCF2014}). Hydrodynamic calculations also show the possibility of
disk formation
at
sub-Keplerian velocities in these systems (Richards \& Ratliff \cite{RR1998}).

Permanent disks are found in systems with $R<\omega_{min}$ . The radial component of the matter velocity in
the
disk is directed towards the accretor and is 10-30~$km/s$.
The
change in the tangential component with
distance from the accretor may differ from the Keplerian one (Etzel et al. \cite{EOS1995}).

The aforementioned observational data and the results of hydrodynamic calculations relate to systems with
the low
mass of accreting components ($M\le6\;M_\odot$) and with a ratio of
donor mass to
accretor mass within the range of 0.2 to 0.3. The formation of Be-stars
in
the early spectral subclass occurs in systems with large component masses. The ratio of
donor mass to
accretor mass varies
widely.
Mass transfer in such systems is non-conservative (Van Rensbergen et al. \cite{RGM11}; Deschamps et al. \cite{DBJ15}). The star
receiving mass increases
in
volume (Benson \cite{B1970}; Kippenhahn \& Meyer-Hofmeister \cite{KMH1977}). The distance between two stars depends on how the system loses mass and angular momentum. So, the formation of gas structures in the Roche lobe of
an
accretor depends on the loss of mass and angular momentum from the system. A quantitative theory of mass and angular momentum losses from a close binary system has not
yet
been developed. The formation of conditions for sub-Kelerian rotation in
an
accretion disk due to
the
loss of mass and angular momentum from the binary system cannot be ruled out. Thus, the possibility of mass accretion with sub-Keplerian velocities of rotation should be considered.

At the very beginning of accretion, when
accretor rotation velocity is
low, the rotation velocity of disk matter decreases in the narrow boundary layer from the maximum value in the disk $\Omega_{max}$ to the value on the star's surface $\Omega_s$ (Paczy{$\acute{\mbox{n}}$}ski \cite{P91}). Turbulence can remove
angular momentum from the boundary layer to
an accretor's
upper layers
at a rate of:
\begin{eqnarray}
\frac{dJ}{dt}&=&\frac{2}{3}R^2(\Omega_{max}-\Omega_{s})\dot M, 
\label{eq001}
\end{eqnarray}
where $J$ - angular momentum of the accretor, $t$ - time, and $\dot M$ - mass accretion rate.

Supersonic shear flow in the boundary layer is a source of acoustic waves. The waves can carry the angular momentum out of the boundary layer both into the accretor's outer part and
the disk's outer part (Dittmann \cite{D2021}, Coleman et al. \cite{CRP2022}). In this case, the amount of angular momentum
coming
from the boundary layer into the accretor is less than the Keplerian one.

In
an earlier
study (Staritsin \cite{St2022}), we
considered
as follows: when at the stage of subcritical rotation,
angular momentum enters the accretor through two channels, namely together with
matter having the same rotation velocity as the accretor's surface and due to turbulence
within
the rate~(1). This is a case of
high efficiency of
angular momentum transfer from the boundary layer to the accretor's upper part. The transfer of
angular momentum in the accretor's interior was carried out by meridional circulation;
turbulence was artificially suppressed.

In the current calculations,
angular momentum transfer in the accretor's interior can be carried out both by meridional circulation and turbulence.
We have studied two variants for the arrival of
angular momentum into the accretor.

In the first variant, to clarify the influence of angular momentum transport by
turbulence in the accretor's interior on
the
spinning up
of the accretor,
we calculated accretion with the same rate of
the
arrival of
angular momentum into the accretor as in
Staritsin (\cite{St2022}). At the stage of subcritical rotation, the parameter $\Omega_{max}$ in
the angular momentum source~(1) is equal to $\alpha\;\Omega^{Kep}$,
where
$\alpha=0.8$; here, $\Omega^{Kep}$ is the Keplerian velocity of the star's surface
at the equator. After the angular velocity of the accretor's surface increases to $\alpha\;\Omega^{Kep}$ value, the arrival of
angular momentum from the boundary layer~(1) stops. The angular velocity of the adding matter
is set equal
to $\alpha\;\Omega^{Kep}$
for
the remainder
of
the
mass exchange.

In the second variant, the case of extremely low efficiency of angular momentum transfer from the boundary layer to the accretor's upper part is considered. The angular momentum's source~(1) in this case is assumed to be zero. As long as the
angular
velocity of the star's surface is less than $\alpha\;\Omega^{Kep}$, the star accretes matter with the same
angular
velocity as that of the star's surface. After the surface
angular
velocity increases to the value of $\alpha\;\Omega^{Kep}$, the
angular
velocity of the adding matter remains equal to $\alpha\;\Omega^{Kep}$. In order to determine the dependence of the
accretor's
rotation state
after the end of
mass exchange on the content of
angular momentum in
adding mass, calculations were carried out at four values of $\alpha$: $0.8$, $0.6$, $0.4$, and $0.2$.

\subsection{Angular momentum transfer in the accretor's interior}
\label{subsect:transfer}

Angular
momentum transfer in the radiative layers of a star is taken into account in the framework of the shellular rotation model (Zahn \cite{Zahn92}). In terms of this model, two mechanisms of angular momentum transfer are considered: meridional circulation and shear turbulence.  The angular momentum transfer is described by the law of conservation of angular momentum (Tassoul \cite{T78}):
\begin{eqnarray}
\label{eq003}
\frac{\partial(\rho\varpi^2\Omega)}{\partial t}+
\mbox{div}(\rho\varpi^2\Omega{\bf u})
=\mbox{div}(\rho\nu_{\mbox{v}}\varpi^2\mbox{grad}\Omega).  \nonumber
\end{eqnarray}
The meridional circulation velocity $\bf u$ is determined from the law of conservation of energy in a stationary form (Maeder \& Zahn \cite{MZ98}):
\begin{eqnarray}
\label{eq004}
\rho T{\bf u}\mbox{grad}s=\rho\varepsilon_n+
\mbox{div}(\chi\mbox{grad}T)-\mbox{div}{\bf F}_h.  \nonumber
\end{eqnarray}
In these equations,
$\rho$ - density,
$\varpi$ - distance to the axis of rotation,
$\Omega$ - angular velocity
$\nu_{\mbox{v}}$ - turbulent viscosity in the vertical direction,
$T$ - temperature,
$s$ - specific entropy,
$\varepsilon_n$ - nuclear energy release rate,
$\chi$ - thermal conductivity,
${\bf F}_h$ - turbulent enthalpy flow in the horizontal direction:
${\bf F}_h=-\nu_h\rho T\partial{s}/\partial{\bf i_\theta}$
and $\nu_h$- turbulent viscosity in the horizontal direction.
The coefficients of turbulent viscosity were determined by Talon and Zahn (\cite{TZ97}), Maeder (\cite{M2003}), and Mathis et al. (\cite{MPZ04}). The convective core rotates solid-state. These equations are solved together with
equations
related to
the structure and evolution of
stars.
We used a set of
programs from
Paczy{$\acute{\mbox{n}}$}ski (\cite{P70})
modified to calculate the evolution of rotating stars
(Staritsin \cite{St99}, \cite{St05}, \cite{St07}, \cite{St09}, \cite{St14}).

   \begin{figure}
   \centering
   \includegraphics[width=\textwidth, angle=0]{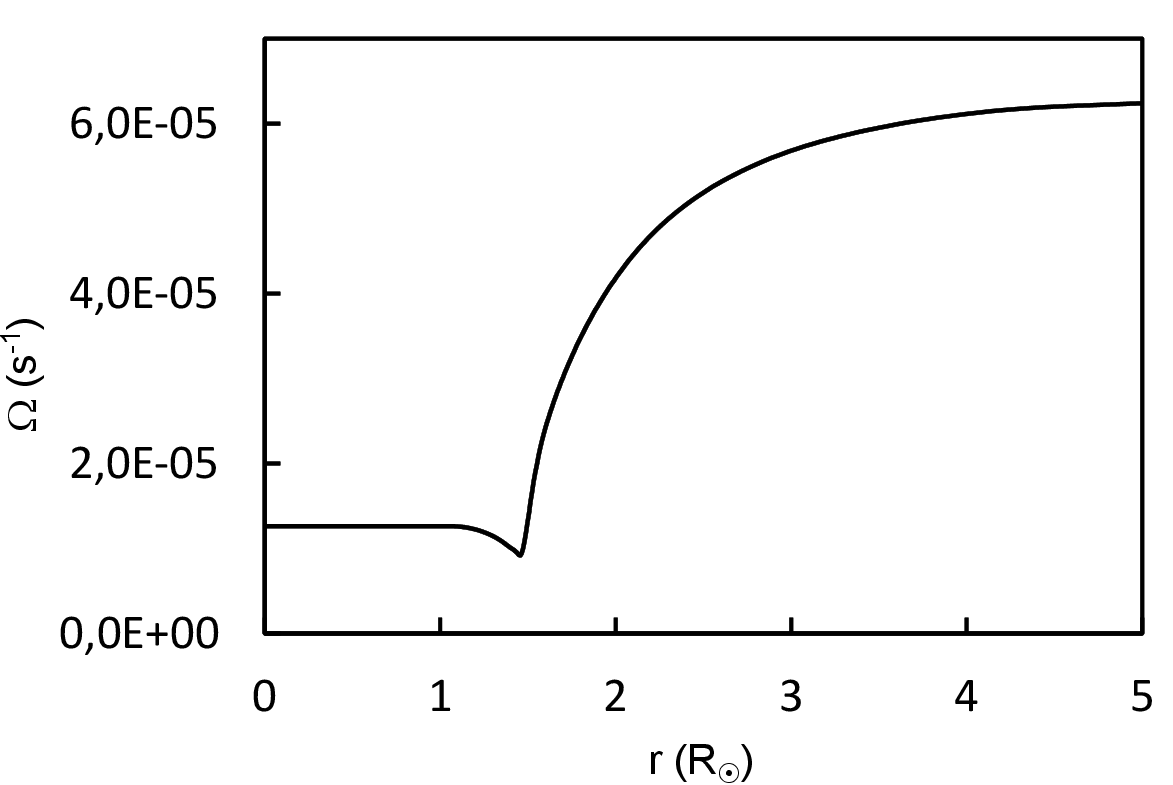}
   \caption{
   Angular velocity at the bottom of the outer cell of the meridional circulation at the beginning of mass exchange.
   }
   \label{Fig1}
   \end{figure}

\section{Calculation results}
\label{sect:calculation}

   \begin{figure}
   \centering
   \includegraphics[width=\textwidth, angle=0]{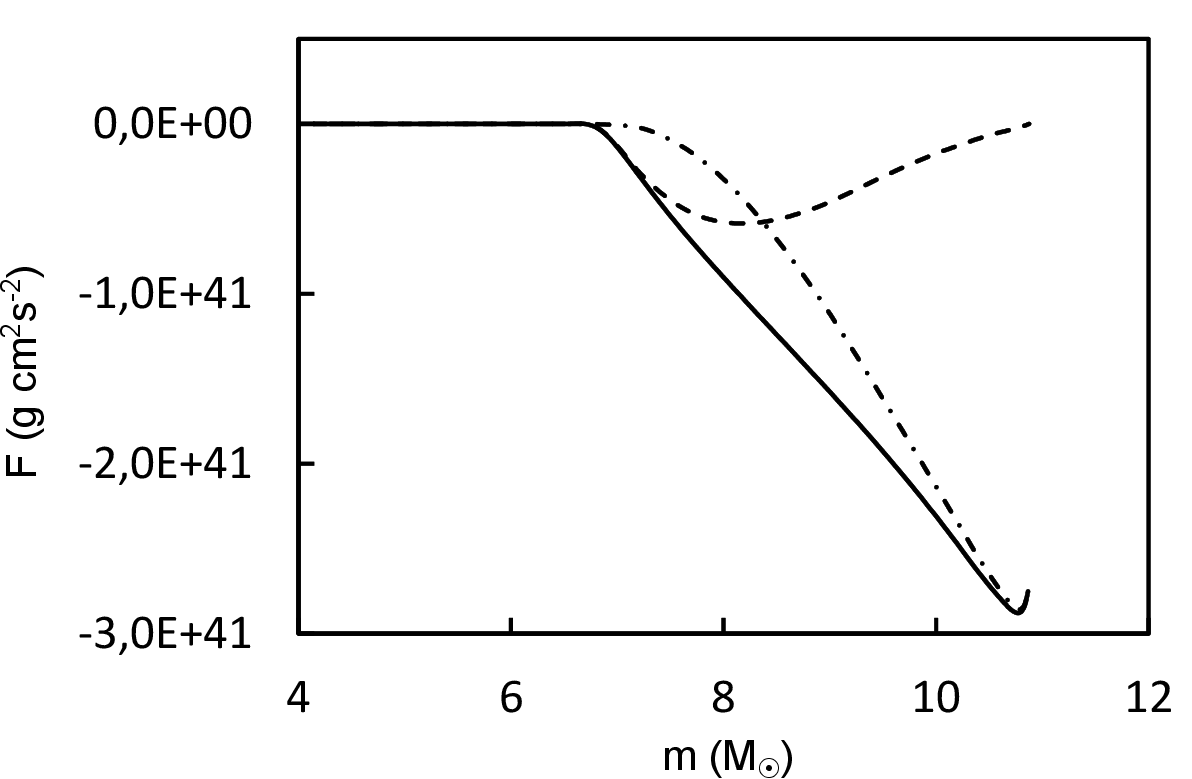}
   \caption{
   Turbulent (dashed-line), advective (dot-and-dashed line), and total (solid line) angular momentum flux in the accretor's interior  at the stage of subcritical rotation.
   }
   \label{Fig2}
   \end{figure}

   \begin{figure}
   \centering
   \includegraphics[width=\textwidth, angle=0]{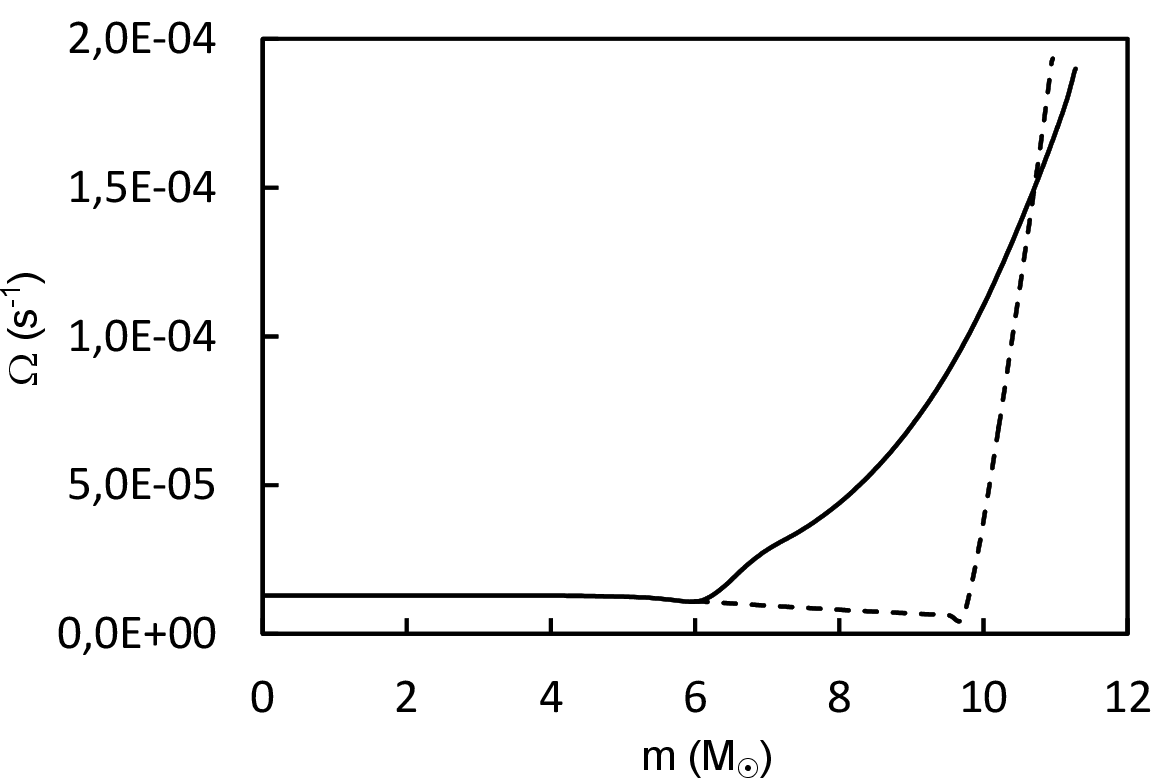}
   \caption{
   Angular velocity when the rotation of the accretor's surface becomes critical and when active turbulence (solid line) and artificially suppressed turbulence are present (Staritsin \cite{St2022})
   (dashed line).
   }
   \label{Fig3}
   \end{figure}

\subsection{Binary system parameters}
\label{subsect:mass}

   \begin{figure}
   \centering
   \includegraphics[width=\textwidth, angle=0]{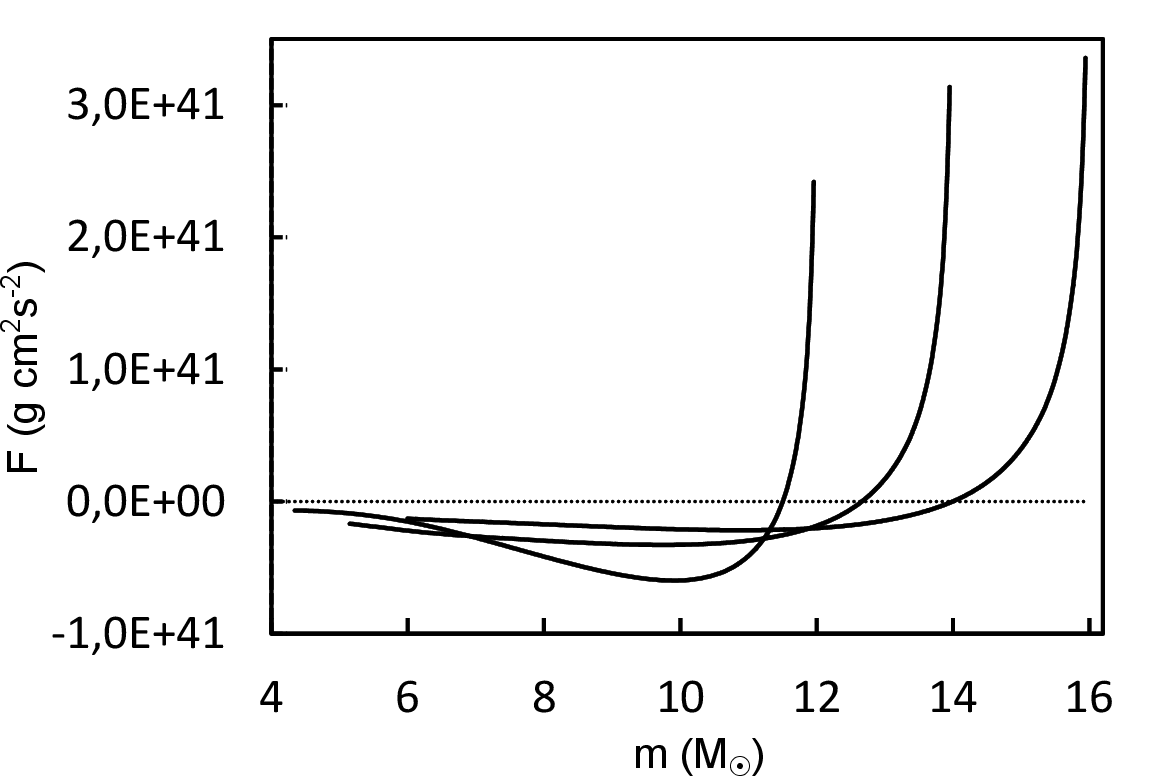}
   \caption{
   Angular momentum flux in the accretor's interior when its mass is equal to 12~$M_\odot$, 14~$M_\odot$, and 16~$M_\odot$.
   }
   \label{Fig4}
   \end{figure}

We
consider
mass exchange in a binary system with the component masses of 13.4~$M_\odot$ and 10.7~$M_\odot$ and the period $P=35^d$ as in Staritsin (\cite{St2022}).
By the beginning of
mass exchange,
star rotation
with a mass of 10.7~$M_\odot$ is synchronized with
orbital motion. The star angular momentum
is
equal
to $1.3\times10^{51}\,g\cdot cm^2s^{-1}$.
A
star with a mass of 13.4~$M_\odot$ loses 10.5~$M_\odot$ for 12,000 years.
After that, the star decouples its Roche lobe and the mass exchange stage ceases.

The second star accretes 5.3~$M_\odot$.
The final mass of the accretor is 16.0~$M_\odot$.
The accretion rate was set constant, equal to the average value of $\sim4.4\times10^{-4}\,M_\odot/year$.
We consider a case when the entropy of the added matter is the same as the surface layers of the second star.
The
thermal timescale of the second star is
longer
than
mass exchange duration. The
star's
reaction
to the increase in mass in this case is well understood (Benson \cite{B1970}; Flannery and Ulrich \cite{FU1977}; Neo et al. \cite{NMNS1977}). The second star is driven out of thermal equilibrium by
mass accretion.
Nuclear
power output in the center of the second star increases, and some of the nuclear energy release is spent on an increase in entropy in the
second star's
central parts.
Gravitational
energy release in the surface layers is added to
nuclear energy release in the center.
The typical
luminosity distribution in the
second star's
interior
is shown in Staritsin (\cite{St2022}) (see Fig. 4).

The remaining part of the mass lost by the first star leaves the system.
The tidal interaction between the two stars is
unable
to synchronize the accreting star with the orbit due to the long period of the system and
the
short accretion timescale. The accretion of matter and angular momentum, as well as
transport processes inside the accretor and in the disk, determine the
accretor's
angular momentum.

\subsection{The case of the high efficiency of angular momentum transfer from the boundary layer to the accretor's upper part}
%
\begin{table}
\begin{center}
\caption[]{ Angular momentum balance.}\label{Tab:publ-works}


 \begin{tabular}{ccccccc}
  \hline\noalign{\smallskip}
No & Case 1 & Case 2 & Case 3 & Case 4 & Case 5 & Case 6 \\
  \hline\noalign{\smallskip}
$J_1$  & 17.2 & 17.2 & 13.3 & 10.3 & 6.9 & 3.7  \\ 
$J_2$  & 11.2 &  9.7 &  6.0 &  3.0 & 0.0 & 0.0  \\
$J_3$  &  5.2 &  5.4 &  5.3 &  5.3 & 5.1 & 2.9  \\
$J_4$  &  0.8 &  2.1 &  2.0 &  2.0 & 1.8 & 0.8  \\
$J_5$  &  6.1 &  7.6 &  7.4 &  7.4 & 7.0 & 3.8  \\
  \noalign{\smallskip}\hline
\end{tabular}
\end{center}
The rows show as follows:\\
($J_1$) the angular momentum that entered the accretor during mass exchange;\\
($J_2$) the angular momentum removed from the accretor during mass exchange;\\
($J_3$) the angular momentum remaining in the accreted mass;\\
($J_4$) the angular momentum transferred to the part of the accretor that made up the star initially;\\
($J_5$) the angular momentum of the accretor after mass exchange,\\
the angular momentum is given in units of $10^{52}\,g\cdot cm^2s^{-1}$.

The columns show the results for the following cases:\\
(Case 1) angular momentum transfer from the boundary layer is considered, turbulence in the accretor's interior is artificially suppressed (Staritsin 2022);\\
(Case 2) angular momentum transfer from the boundary layer is considered, turbulence in the accretor interior is considered;\\
(Case 3), (Case 4), (Case 5), and (Case 6) angular momentum transfer from the boundary layer is not considered, turbulence in the accretor interior is present, and $\alpha$ is equal to 0.8, 0.6, 0.4, and 0.2 respectively.
\end{table}

With the beginning of mass exchange, a circulation cell is formed in the subsurface layer of the accretor, in which the circulation carries the incoming angular momentum
downwards.
The cell consists of accreted layers and the swirled layers of the accretor located below. In the
cell's
upper part,
angular velocity has an almost constant value, but near the bottom of the cell,
it
sharply reduces to
the
initial value (Fig.~1). Therefore, in the lower part of the cell, the contribution of turbulence to
angular momentum transfer is greater and exceeds the contribution of meridional circulation (Fig.~2). The bottom of the cell goes down into the star faster than
when turbulence is artificially suppressed. The angular momentum entering the accretor is distributed over a larger mass of matter than in the case of suppressed turbulence. The rotation of the accretor's surface becomes critical when its mass increases to 11.3~$M_\odot$ (in the case of suppressed turbulence - up to 11.0~$M_\odot$ (Staritsin \cite{St2022})). The distribution of angular velocity in the accretor's interior at this moment is shown in Fig.~3.

At the stage of critical rotation, the mass of the accretor increases by another 4.7~$M_\odot$. Another circulation cell is formed in the accreted matter.
In this cell, the circulation transfers the some part of the angular momentum that came along with the accreted mass to the surface of the accretor (Fig.~4).
It is assumed that this part of the angular momentum is removed from the accretor by the accretion disk (Paczy{$\acute{\mbox{n}}$}ski \cite{P91}; Bisnovatyi-Kogan \cite{BK93}).
As a result of a decrease in
angular momentum, the accreted layers are contracted. The velocity of their rotation
is permanently lower
than the Keplerian velocity.

In the circulation cell
formed at the beginning of
mass exchange, the transfer of
angular momentum inside the accretor continues. The mass of the matter in this cell increases as the upper boundary of the cell moves up along the accretor mass, and the bottom of the cell moves down. The bottom of the cell goes down to the convective core when the accretor mass increases to 11.9~$M_\odot$ (in the case of suppressed turbulence - up to 15~$M_\odot$ (Staritsin \cite{St2022})). The role of turbulence lies in the
rapid lowering of the bottom of the circulation cell, in which the circulation carries
angular momentum into the star's interior.

The amount of angular momentum removed from the accretor during
mass exchange depends slightly on
processes of angular momentum transfer within the accretor (Fig.~5).
When
turbulence
is present,
the amount of angular momentum transferred to the accretor's inner layers increases, and the amount of angular momentum carried to the accretor's surface decreases compared to the case of suppressed turbulence (Table 1).
The
angular momentum brought into the accretor during
mass exchange is $1.72\times10^{53}\,g\cdot cm^2s^{-1}$. 12\% of this value enters the inner layers that made up the accretor initially, 31\% remains in the accreted mass, and 57\% is carried to the accretor's surface and is removed
by
the
disk. In the case of suppressed turbulence, the corresponding values are 5\%, 30\%, and 65\% (Staritsin \cite{St2022}). After the end of
mass exchange, the accretor's angular momentum is greater when turbulence
is present
(Table 1).

\subsection{The case of extremely low efficiency of angular momentum transfer from the boundary layer to the accretor's upper part}

   \begin{figure}
   \centering
   \includegraphics[width=\textwidth, angle=0]{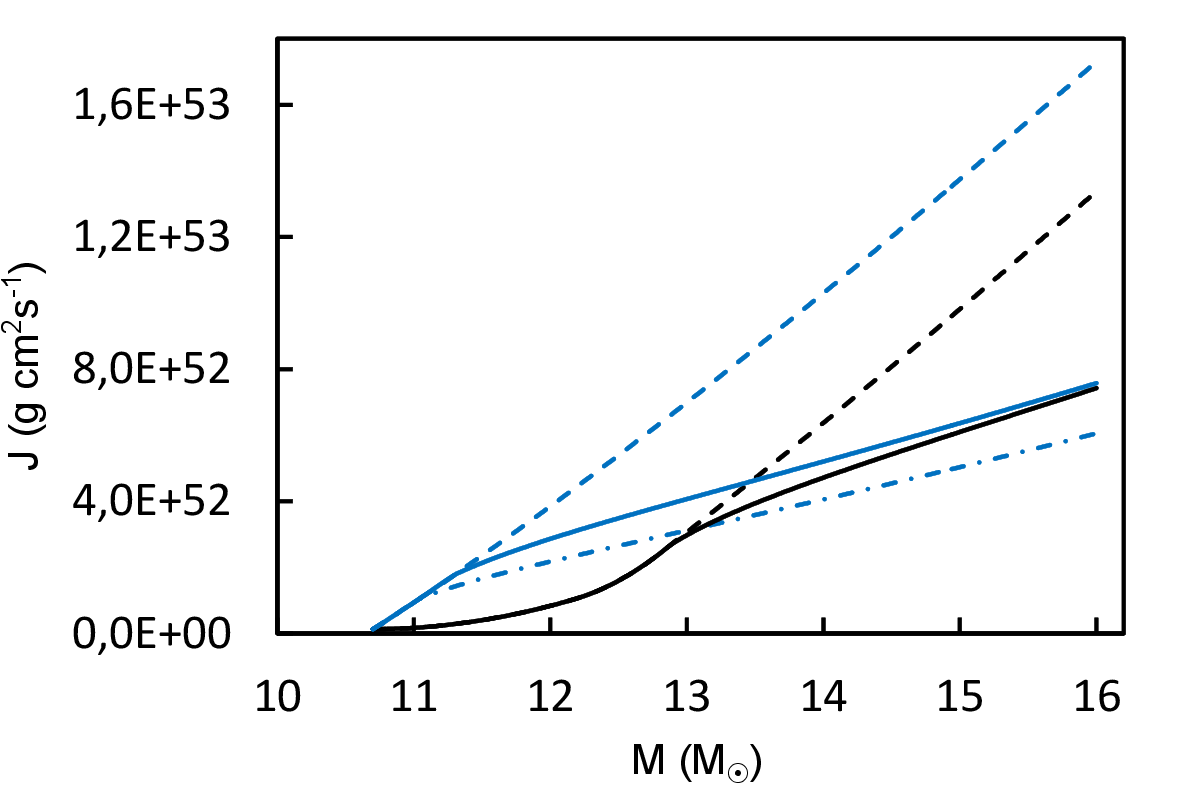}
   \caption{
   The angular momentum amount entering the accretor (dashed line), the angular momentum of the accretor (solid line) depending on its mass at $\alpha=0.8$ when the source of the angular momentum~(1) is considered (blue color) and is not taken into account (black color).The case of artificially suppressed turbulence (Staritsin \cite{St2022}) is also
   shown (dot-and-dashed line).
   }
   \label{Fig5}
   \end{figure}

   \begin{figure}
   \centering
   \includegraphics[width=\textwidth, angle=0]{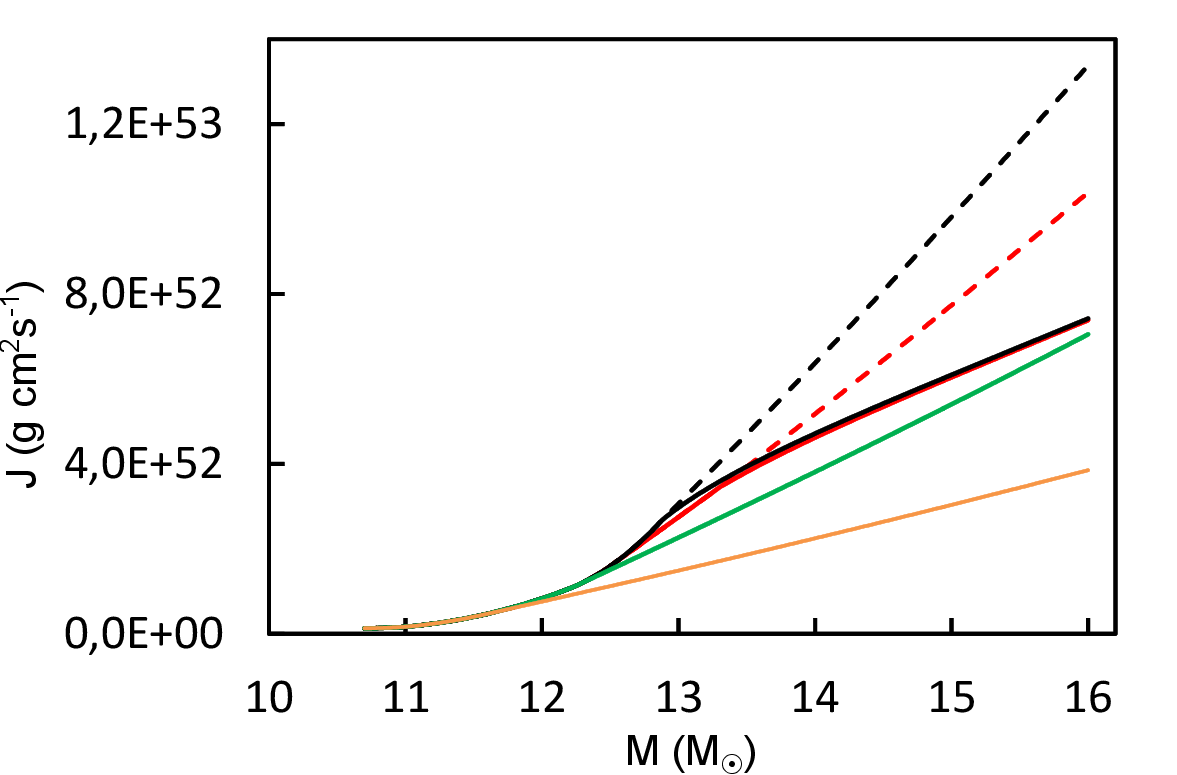}
   \caption{
   The angular momentum amount
   entering
   the accretor (dashed line), the angular momentum of the accretor (solid line) depending on its mass at $\alpha=0.8$ (black), $\alpha=0.6$ (red), $\alpha=0.4$ (green), and $\alpha=0.2$ (orange) when the source of angular momentum~(1) is not considered.
   At $\alpha=0.4$ and $\alpha=0.2$, the accretor retains the entire angular momentum obtained with the accreted mass, so the dashed lines coincide with the solid ones.
   }
   \label{Fig6}
   \end{figure}

At the beginning of
mass exchange, the rotation velocity of the incoming mass and the accretor's surface coincide. The rate of angular momentum arrival into the accretor is significantly less than
when turbulence and/or waves transfer
angular momentum from the boundary layer to the accretor's outer part. Due to the low rate at which
angular momentum enters the accretor, the
accretor's
angular momentum
increases slowly at the beginning of accretion (Fig.~5).

The general picture of
angular momentum transfer in the accretor's interior at $\alpha$ equal to 0.8 and 0.6 is the same as when
angular momentum goes from the boundary layer to the accretor's upper part.The difference is that
the total amount of angular momentum that has entered the accretor during
mass exchange decreases (Table 1).
Reasons for the decrease are associated with the absence of a source~(1) and
a
decrease in
parameter
$\alpha$.
However, with $\alpha$ equal to 0.8 and 0.6, the
accretor surface's
rotation velocity
increases to a critical value. This occurs when
accretor mass increases to 12.9~$M_\odot$ and 13.3~$M_\odot$, respectively (Fig.~6). In these cases, a circulation cell is formed in the accretor's outer layer, in which the circulation transfers part of the angular momentum of the accreted layers to the accretor's surface. The amount of angular momentum removed from the accreted layers and lost by the accretor in these cases is less than in calculations with
a
source~(1) (Table 1). The state of
accretor rotation once mass exchange finishes is approximately the same as when the angular momentum's arrival from the boundary layer to the accretor's upper part was considered.
A decrease in angular momentum entering the accretor only results in
decreases
in the angular momentum
taken out of the accretor at the stage of accretion
during
accretor critical rotation.

At $\alpha$ equal to 0.4 and 0.2, a smaller
amount of
angular momentum enters the accretor (Table 1). The rotation velocity of the accretor's surface remains subcritical throughout the entire mass exchange stage;
at $\alpha$ equal to 0.4,
it
approaches the critical value by the end of this stage. In both cases, at the beginning of mass exchange, a circulation cell is formed in the accretor's subsurface layer, in which the angular momentum of the accreted matter is transferred inside the accretor. The bottom of the cell goes down to the convective core when the mass of the accretor increases to 13.1~$M_\odot$ at $\alpha$ equal to 0.4 and up to 13.9~$M_\odot$ at $\alpha$ equal to 0.2. In both cases, the angular momentum of the accreted mass is transferred inside the star throughout the
mass exchange stage. The accretor retains
all
the
angular momentum received with the accreted mass (Fig.~6). Once
mass exchange finishes, the angular momentum of the accretor at $\alpha$ equal to 0.4 is little less than at $\alpha$ equal to 0.6 and 0.8, and at $\alpha$ equal to 0.2 is significantly less (Table 1).

\subsection{Accretor rotation state after mass exchange}

The distribution of the angular velocity of rotation in the accretor's interior immediately after the end of
mass exchange is shown in Fig. 7. In all cases,
angular velocity decreases rapidly in a layer of variable chemical composition located between the chemically homogeneous part of the radiative envelope and the convective core. A similar jump is formed in cases where
angular momentum enters the accretor in a short time - in the donor's thermal timescale or faster (Staritsin \cite{St21}). The thermal timescale of the accretor is
longer
than that of the donor in the cases considered in Staritsin (\cite{St21}, \cite{St2022}). After the end of
mass exchange, the jump gradually decreases and disappears during the thermal timescale of the accretor (see, for example, Fig.~3 in Staritsin (2021)).

The angular velocity
in the accretor's interior after mass exchange when $\alpha$
is
equal to 0.8 and $\alpha$
is
equal to 0.6 almost does not depend on what
the content of the angular momentum
was
in the adding mass and on whether
angular momentum is transferred from the boundary layer to the accretor's upper layer
(Fig.~7). In these cases, the accretor's angular momentum is almost
equal to $\sim7.5\times10^{52}\,g\cdot cm^2s^{-1}$ (Table 1) after the mass exchange.  An isolated star with a mass of 16~$M_\odot$ has a critical rotation
throughout the
stage of hydrogen burning in the core with this angular momentum value (Staritsin \cite{St07}). Consequently, due to the exchange of mass, the accretor receives a rotation typical for Be-stars.

At $\alpha$ equal to 0.4, the accretor receives almost the same angular momentum with accreted mass as the angular momentum that remains in the accretor when $\alpha$ is 0.8 and $\alpha$ is 0.6 (Table 1). Therefore, at $\alpha$ equal to 0.4, the accretor also has a rotation typical for Be-stars.

At $\alpha$ equal to 0.2, the accreted mass brings a much
lower
angular momentum (Table 1). The angular velocity in the accretor's interior is
lower
than in other cases (Fig.~7). The rotation of the accretor's surface immediately after
mass exchange in this case is
lower
than that of Be-type stars. In an isolated star with the same mass and angular momentum as
the
accretor, the removal of
angular momentum from the inner layers to the outside occurs intensively at the stage of hydrogen burning in the core (Staritsin \cite{St07}, \cite{St09}). The angular velocity of the star's surface, expressed in
Keplerian angular velocity, increases;
at the last steps of this stage, the star acquires a rotation typical for Be-stars of the early spectral subclass. If
tidal interaction is
low, then even in this case the accretor
can
obtain the characteristics of
a
Be-star after the end of mass exchange, but only after a long period of time
on
the order of
part of the hydrogen burning stage in the core.

\section{Conclusions}

   \begin{figure}
   \centering
   \includegraphics[width=\textwidth, angle=0]{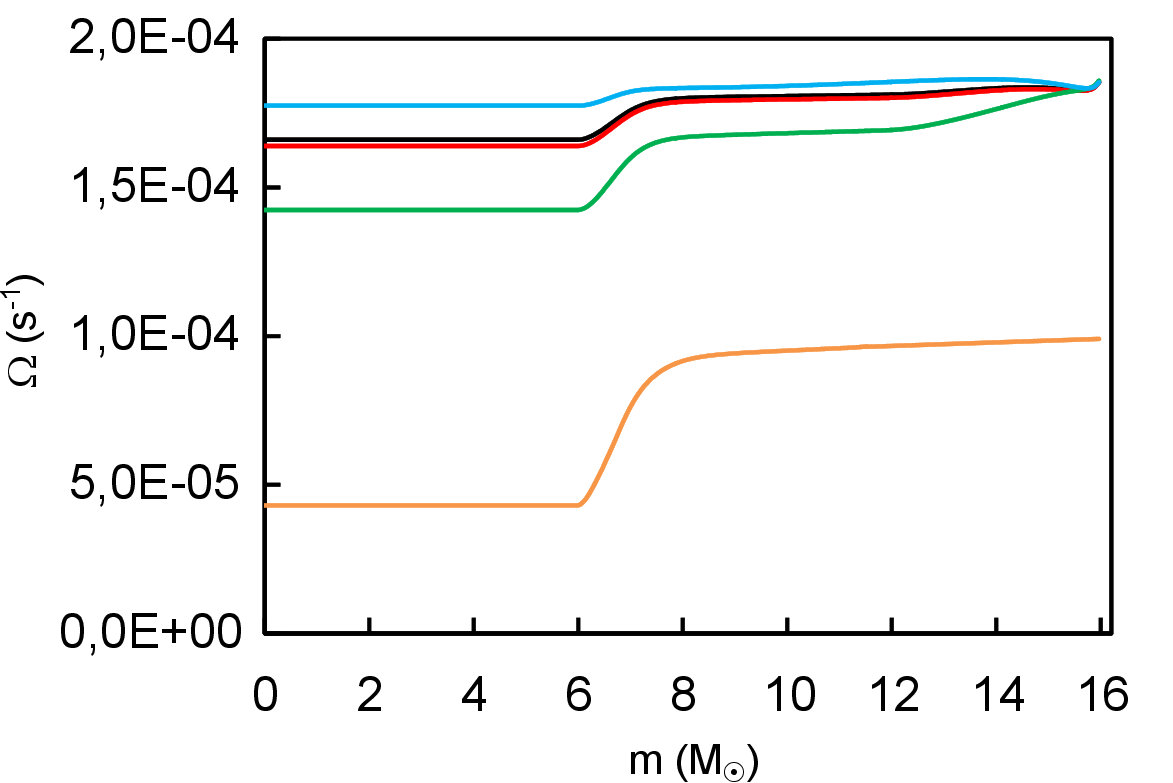}
   \caption{
   The distribution of
   angular velocity in the accretor's interior after the end of mass exchange in cases when the source of angular momentum~(1) is considered (blue) and not considered at $\alpha=0.8$ (black), $\alpha=0.6$ (red), $\alpha=0.4$ (green), and $\alpha=0.2$ (orange).
   }
   \label{Fig7}
   \end{figure}

Meridional circulation is a flexible mechanism for the transfer of angular momentum in
the
stellar interior of a rotating star. The direction and rate of angular momentum transfer by circulation may vary widely at the stage of mass accretion, depending on the star rotation state and the rate of angular momentum arrival along with the accreted mass and waves and/or due to turbulence. The two main circulation cells are formed due to
the
accretion of mass and angular momentum. In the cell, which is formed at the stage of subcritical rotation of the accretor, circulation transfers the angular momentum inside the accretor. Only in the lower part of this cell
does
turbulence
make
the main contribution to the transfer of angular momentum. Due to turbulence, the cell bottom quickly goes
downwards
into
the accretor's interior. In a cell
formed at the stage of critical rotation, the circulation transfers part of the angular momentum of the accreted mass to the surface of the star;
the more the content of the angular momentum in the entering matter is,
the greater this part.

We have considered the case of mass exchange in a binary system, when half of the mass lost by the donor falls
into
the accretor.
If the angular momentum of the mass falling
into
the accretor exceeds half
the Keplerian value at the boundary of the accretor, the state of rotation of the accretor after the end of
mass exchange does not depend on the angular momentum entering the accretor. In other words, processes that could reduce the angular momentum of the
matter located around the accretor to no more than half the Keplerian value do not affect the angular momentum and the state of rotation that the accretor receives
by the end of the mass exchange stage. These processes impact on the amount of angular momentum removed by circulation from the accreted mass to the accretor's surface
and removed further from the accretor by a disk or other processes.

In the considered system with the initial
component
masses
of 13.4~$M_\odot$ and 10.7~$M_\odot$, the accretor
has
a rotation typical for Be-stars immediately after the end of mass exchange, if
during mass exchange the angular momentum of the mass added to the accretor exceeded 40\% of the Keplerian value.

\begin{acknowledgements}
This work was supported by the Ministry of Science and Education, FEUZ-2023-0019
\end{acknowledgements}

%

\label{lastpage}

\end{document}